\documentclass[10pt,a4paper,twoside]{article}
\usepackage{epsfig}
\usepackage{baltlat6}
\usepackage{wrapfig}
\usepackage{array}
\usepackage{dcolumn}
\usepackage{longtable}

\begin{document}
\ \
\vspace{-0.5mm}

\setcounter{page}{157}
\vspace{-2mm}

\titlehead{Baltic Astronomy, vol.\,19, 157--167, 2010}

\titleb{CHEMICAL COMPOSITION OF THE RS CVn-TYPE STAR\\
29~DRACONIS}

\begin{authorl}
\authorb{G.~Barisevi\v{c}ius,}{1}
\authorb{G.~Tautvai\v{s}ien\.{e},}{1}
\authorb{S.~Berdyugina,}{2}
\authorb{Y.~Chorniy}{1} and
\authorb{I.~Ilyin}{3}
\end{authorl}

\begin{addressl}
\addressb{1}{Institute of Theoretical Physics and Astronomy, Vilnius University,\\
Go\v{s}tauto 12, Vilnius, LT-01108, Lithuania}

\addressb{2}{Kiepenheuer Institut f\"ur Sonnenphysik, Sch\"oneckstr. 6,
D-79104 Freiburg,\\ Germany}

\addressb{3}{Astrophysical Institute Potsdam, An der Sternwarte 16,
Potsdam D-14482,\\ Germany}
\end{addressl}

\submitb{Received 2010 October 2; accepted 2010 November 5}

\begin{summary} Photospheric parameters and chemical composition are
determined for the single-lined chromospherically active RS~CVn-type
star 29 Draconis (HD 160538).  From the high resolution spectra obtained
on the Nordic Optical Telescope, abundances of 22 chemical elements,
including the key elements such as $^{12}{\rm C}$, $^{13}{\rm C}$, N and
O, were investigated.  The differential line analysis with the MARCS
model atmospheres gives $T_{\rm eff}=4720$~K, log~$g=2.5$, ${\rm
[Fe/H]}=-0.20$, ${\rm [C/Fe]}=-0.14$, ${\rm [N/Fe]}=0.08$, ${\rm
[O/Fe]}=-0.04$, ${\rm C/N}=2.40$, $^{12}$C/$^{13}$C = 16.  The low value
of the $^{12}$C/$^{13}$C ratio gives a hint that extra mixing processes
in low-mass chromospherically active stars may start earlier than
the theory of stellar evolution predicts.  \end{summary}


\resthead{Chemical composition of the RS CVn-type star 29 Draconis}{ G.~
Barisevi\v{c}ius, G.~Tautvai\v{s}ien\.{e}, S.~Berdyugina et al.}

\sectionb{1}{INTRODUCTION}

In this work we continue a detailed study of photospheric abundances in
RS CVn stars (Tautvai\v{s}ien\.{e} et al.\ 2010, hereafter Paper~I) with
the aim to get new data on chemical evolution of the photospheric and
coronal abundance patterns and mixing processes in chromospherically
active stars.  Our aim is to investigate abundances of more than 20
chemical elements, including $^{12}$C, $^{13}$C, N, O and other mixing
sensitive species.  We plan to investigate correlations between the
abundance alterations of chemical elements in star atmospheres and their
physical macro parameters, such as the speed of rotation and the
magnetic field.

Here we present the results of the analysis for the RS CVn binary 29
Draconis (HD 160538) consisting of a K0\,III star ($V \approx 6.64$~mag)
and a white dwarf.  Since the primary dominates the spectrum, the
photospheric abundances can be relatively well determined, and thus the
system is a good candidate for studying the coronal-to-photospheric
abundance patterns.

Hall et al.\ (1982) were the first to suggest that 29~Draconis is a
RS~CVn binary.  Its late spectral type (K0\,III according to Bidelman
1954 or K2\,III according to Roman 1955), the Ca\,{\sc ii} H \& K
emission in its spectrum (Bidelman 1954), radial velocity values by Abt
\& Biggs (1972) and photometric variability reaching $\Delta V =
0.12$~mag, as well as a rotation period of 31.5~days found by Hall
et al., allowed the authors to conclude that 29~Dra is a RS~CVn binary.

Ultraviolet spectra of 29~Dra were observed by Fekel \& Simon (1985).
It was found that the absorption lines are slightly broadened by stellar
rotation with $v\,{\rm sin}\,i = 8\pm 2$~km\,s$^{-1}$.  The H$\alpha$
absorption line was weak, as if filled in by emission, and variable.  In
five observations obtained over a period of four months, the equivalent
width ranged from 0.1 to 0.8~\AA.  Mg\,{\sc ii} $h$ and $k$ lines were
in emission.  The study by Fekel \& Simon has revealed a white dwarf
companion of 29~Dra, whose ultraviolet spectrum is matched best to the
energy distribution of a $T_{\rm eff}=30\,000$~K and log\,$g=8$ model
atmosphere.  The mass and radius of the white dwarf were found to be
0.55~$M_{\odot}$ and 0.012~$R_{\odot}$, respectively.  A radius of the
primary component was evaluated to be more than 5~$R_{\odot}$.  Fekel et
al.\ (1993) derived $v\,{\rm sin}\,i = 7.2$~km\,s$^{-1}$, the orbital
period 903.8 days and the parameters for the companion star.  They
suspected that 29\,Dra can be a barium-like star.

However, high-resolution spectral analyses by Berdyugina (1994) and
Za\v{c}s et al.\ (1997) have ruled out the hypothesis that 29\,Dra is a
barium-like star.

In further investigations of 29~Dra, Fekel (1997) determined a more
accurate value of the radius of the primary component, 4.2~$R_{\odot}$,
and the projected rotational velocity $v\,{\rm sin}\,i =
6.7$~km\,s$^{-1}$, which was used in our work.

A long-term photometry of 29~Dra, which lasted about 28 years, was
recently reported by Zboril \& Messina (2009).  They found that this
star has three magnetic cycles (20.3, 11.1 and 7.6 yr).  Modeling of the
light curve indicated that the spots cover from 4 to 10\% of total
stellar surface.  The photometric period was found to be variable in the
range 26.4--31.8 d, suggesting differential rotation.

In our recent investigation of $\lambda$~And (Paper~I), we found a low
ratio of $^{12}$C/$^{13}$C = 14, which is lower than is predicted by the
first dredge-up theory for a first-ascent giant with the parameters of
$\lambda$~And.  Masses and luminosities of 29~Dra and $\lambda$~And are
very similar.  In this work, it was interesting to check a hint that
extra mixing processes in low-mass chromospherically active fast
rotating stars may start earlier than in non-active stars, i.e., below
the bump of the luminosity function of red giant branch.

\sectionb{2}{OBSERVATIONS AND THE METHOD OF ANALYSIS}

The spectra of 29~Dra were obtained in 1999 August on the 2.56~m Nordic
Optical Telescope using the SOFIN echelle spectrograph with the optical
camera, which provided a spectral resolving power of $R \approx
80\,000$, for 26 spectral orders, each of $\sim 40$~\AA, in the spectral
region from 5000 to 8300~\AA.  The details of spectral reductions were
presented in Paper~I.

In the spectra we selected 148 atomic lines for the measurement of
equivalent widths (Table~1) and 17 lines for the comparison with
synthetic spectra.  The spectra were analyzed using a differential model
atmosphere technique described in Paper~I.  Here we present
only some details.

\begin{table}[!t]
{
\footnotesize
\noindent
\extrarowheight=-.5pt
\tabcolsep=2pt
\begin{longtable}{lcD..{3.3}|lcD..{3.3}|lcD..{3.3}}
\multicolumn{9}{c}{\parbox[c]{120mm}{\baselineskip=4pt
{\smallbf\ \ Table 1.}{\small\ \ Measured equivalent  widths of lines,
{\it EW}, in the spectrum of 29 Dra.\lstrut}}}\\
\tablerule
\noalign{\vskip0.5mm}
\multicolumn{1}{l}{Element} &
\multicolumn{1}{c}{$\lambda$\,(\AA)} &
\multicolumn{1}{c|}{{\it EW}\,(m\AA)} &
\multicolumn{1}{l}{Element} &
\multicolumn{1}{c}{$\lambda$\,(\AA)} &
\multicolumn{1}{c|}{{\it EW}\,(m\AA)} &
\multicolumn{1}{l}{Element} &
\multicolumn{1}{c}{$\lambda$\,(\AA)} &
\multicolumn{1}{c}{{\it EW}\,(m\AA)}\\
\noalign{\vspace{0.5mm}}
\tablerule
\noalign{\vspace{0.5mm}}
\endfirsthead
\multicolumn{9}{l}{{\smallbf\ \ Table 1.}{\small\ Continued\lstrut}}\\
\tablerule
\noalign{\vskip0.5mm}
\multicolumn{1}{l}{Element} &
\multicolumn{1}{c}{$\lambda$\,(\AA)} &
\multicolumn{1}{c|}{{\it EW}\,(m\AA)} &
\multicolumn{1}{l}{Element} &
\multicolumn{1}{c}{$\lambda$\,(\AA)} &
\multicolumn{1}{c|}{{\it EW}\,(m\AA)} &
\multicolumn{1}{l}{Element} &
\multicolumn{1}{c}{$\lambda$\,(\AA)} &
\multicolumn{1}{c}{{\it EW}\,(m\AA)}\\
\noalign{\vspace{0.5mm}}
\tablerule
\noalign{\vspace{0.5mm}}
\endhead
\endfoot
~~Si\,{\sc i}	 &    5517.55 &   15.1 & 	 &    6266.30 &   47.6 & 	 &    6786.86 &   49.0 \\
	 &    5645.60 &   55.0 & 	 &    6274.66 &   75.4 & 	 &    6793.27 &   30.9 \\
	 &    5665.55 &   55.1 & 	 &    6285.16 &   83.1 & 	 &    6839.83 &   83.2 \\
	 &    5793.08 &   50.5 & 	 &    6292.82 &   86.8 & 	 &    6842.69 &   59.6 \\
	 &    5948.54 &   92.1 &  &  &  & 	 &    6843.65 &   86.0 \\
	 &    6131.85 &   28.8 & ~~Cr\,{\sc i}	 &    5712.78 &   49.6 & 	 &    6851.64 &   34.4 \\
	 &    7003.57 &   51.3 & 	 &    5783.87 &   75.4 & 	 &    6857.25 &   45.8 \\
 &  &  & 	 &    5784.97 &   58.3 & 	 &    6858.15 &   73.9 \\
~~Ca\,{\sc i}	 &    5260.38 &   76.3 & 	 &    5787.92 &   82.8 & 	 &    6862.49 &   50.5 \\
	 &    5867.57 &   60.0 & 	 &    6661.08 &   27.4 & 	 &    7461.53 &   75.0 \\
	 &    6455.60 &  107.7 & 	 &    6979.80 &   71.3 &  &  &  \\
	 &    6798.47 &   28.1 & 	 &    6980.91 &   29.6 & ~~Fe\,{\sc ii}	 &    5132.68 &   37.4 \\
 &  &  &  &  &  & 	 &    5264.81 &   49.2 \\
~~Sc\,{\sc ii}	 &    5526.81 &  111.8 & ~~Fe\,{\sc i}	 &    5395.22 &   39.9 & 	 &    6113.33 &   17.5 \\
	 &    5640.98 &   77.4 & 	 &    5406.78 &   59.7 & 	 &    6369.46 &   25.0 \\
	 &    5667.14 &   62.5 & 	 &    5522.45 &   67.4 & 	 &    6456.39 &   64.7 \\
	 &    6279.75 &   57.3 & 	 &    5577.03 &   21.0 & 	 &    7711.72 &   44.8 \\
	 &    6300.69 &   15.9 & 	 &    5579.35 &   24.4 &  &  &  \\
 &  &  & 	 &    5607.67 &   36.7 & ~~Co\,{\sc i}	 &    5530.78 &   71.2 \\
~~Ti\,{\sc i}	 &    5648.58 &   45.4 & 	 &    5608.98 &   29.7 & 	 &    5590.71 &   60.6 \\
	 &    5662.16 &   69.3 & 	 &    5651.48 &   35.4 & 	 &    5647.23 &   51.7 \\
	 &    5716.45 &   32.5 & 	 &    5652.33 &   48.7 & 	 &    6117.00 &   41.8 \\
	 &    5739.48 &   42.5 & 	 &    5653.86 &   62.3 & 	 &    6188.98 &   59.1 \\
	 &    5880.27 &   54.6 & 	 &    5679.03 &   81.4 & 	 &    6455.00 &   37.7 \\
	 &    5899.30 &  104.6 & 	 &    5720.90 &   31.1 & 	 &    6595.86 &   17.9 \\
	 &    5903.31 &   46.5 & 	 &    5732.30 &   24.0 & 	 &    6678.82 &   38.8 \\
	 &    5941.76 &   73.0 & 	 &    5738.24 &   31.0 &  &  &  \\
	 &    5953.17 &   88.1 & 	 &    5741.86 &   54.0 & ~~Ni\,{\sc i}	 &    5578.73 &   98.1 \\
	 &    5965.83 &   84.5 & 	 &    5784.67 &   59.9 & 	 &    5587.87 &  100.2 \\
	 &    6064.63 &   62.4 & 	 &    5793.92 &   59.5 & 	 &    5589.37 &   48.4 \\
	 &    6098.66 &   23.3 & 	 &    5806.73 &   75.0 & 	 &    5593.75 &   61.0 \\
	 &    6121.00 &   29.3 & 	 &    5807.79 &   28.1 & 	 &    5643.09 &   27.8 \\
	 &    6126.22 &   88.6 & 	 &    5809.22 &   81.5 & 	 &    5748.35 &   77.6 \\
	 &    6220.49 &   45.4 & 	 &    5811.92 &   25.9 & 	 &    5805.22 &   56.7 \\
	 &    6303.77 &   54.5 & 	 &    5814.82 &   44.0 & 	 &    6053.68 &   32.0 \\
	 &    6599.11 &   65.7 & 	 &    6027.06 &   89.0 & 	 &    6108.12 &  113.8 \\
	 &    6861.45 &   39.7 & 	 &    6034.04 &   19.8 & 	 &    6111.08 &   50.7 \\
 &  &  & 	 &    6035.35 &   15.2 & 	 &    6128.98 &   71.5 \\
~~V\,{\sc i}	 &    5604.96 &   34.8 & 	 &    6054.07 &   22.1 & 	 &    6130.14 &   33.8 \\
	 &    5646.11 &   45.3 & 	 &    6056.01 &   90.0 & 	 &    6204.60 &   38.4 \\
	 &    5657.45 &   52.6 & 	 &    6098.25 &   31.6 & 	 &    6378.25 &   49.4 \\
	 &    5668.37 &   52.9 & 	 &    6105.13 &   22.8 & 	 &    6482.80 &   85.6 \\
	 &    5670.86 &   85.3 & 	 &    6120.24 &   45.8 & 	 &    6586.32 &   91.3 \\
	 &    5727.66 &   68.1 & 	 &    6187.99 &   77.1 & 	 &    6598.60 &   38.3 \\
	 &    5737.07 &   76.4 & 	 &    6200.32 &  123.8 & 	 &    6635.13 &   36.2 \\
	 &    5743.43 &   59.1 & 	 &    6226.74 &   56.8 & 	 &    6767.78 &  140.1 \\
	 &    6039.74 &   72.9 & 	 &    6229.23 &   83.6 & 	 &    6772.32 &   75.3 \\
	 &    6058.18 &   36.1 & 	 &    6270.23 &   95.5 & 	 &    6842.03 &   48.0 \\
	 &    6111.65 &   76.3 & 	 &    6380.75 &   79.8 & 	 &    7001.55 &   46.9 \\
	 &    6119.53 &   87.2 & 	 &    6392.54 &   60.0 & 	 &    7062.97 &   50.9 \\
	 &    6135.37 &   70.6 & 	 &    6574.21 &   99.0 & 	 &    7715.59 &   74.9 \\
	 &    6224.50 &   68.8 & 	 &    6581.21 &   78.8 &  &  &  \\
	 &    6233.19 &   65.6 & 	 &    6646.97 &   43.4 &  &  &  \\
\tablerule
\end{longtable}
}
\end{table}

\subsectionb{2.1}{Atmospheric parameters}

Initially, the effective temperature, $T_{\rm eff}$, of 29~Dra was
derived and averaged from the intrinsic color indices $(B-V)_0$ and
$(b-y)_0$ using the corrected calibrations by Alonso et al.\ (1999).
The values
of color indices $B-V=1.043$ and $b-y=0.664$ were taken from van Leeuwen
et al.\ (2007) and Hauck \& Mermilliod (1998), respectively.  A small
dereddening correction of $E_{B-V}=0.02$, estimated using the Hakkila et
al.\ (1997) software, was taken into account.

The agreement between the temperatures deduced from the two color
indices was quite good, with the difference 70~K only.  No obvious trend
of the Fe\,{\sc i} abundances with the excitation potential was found
(Figure~1).

The surface gravity log\,$g$ was found by adjusting the model gravity to
yield the same iron abundance from the Fe\,{\sc i} and Fe\,~{\sc ii}
lines.  The microturbulent velocity $v_{\rm t}$ value corresponding to a
minimal line-to-line Fe\,{\sc i} abundance scattering was chosen as a
correct value.  Consequently, [Fe/H] values do not depend on the
equivalent widths of lines (Figure~2).

\vspace{3mm}
\vbox{
\centerline{\psfig{figure=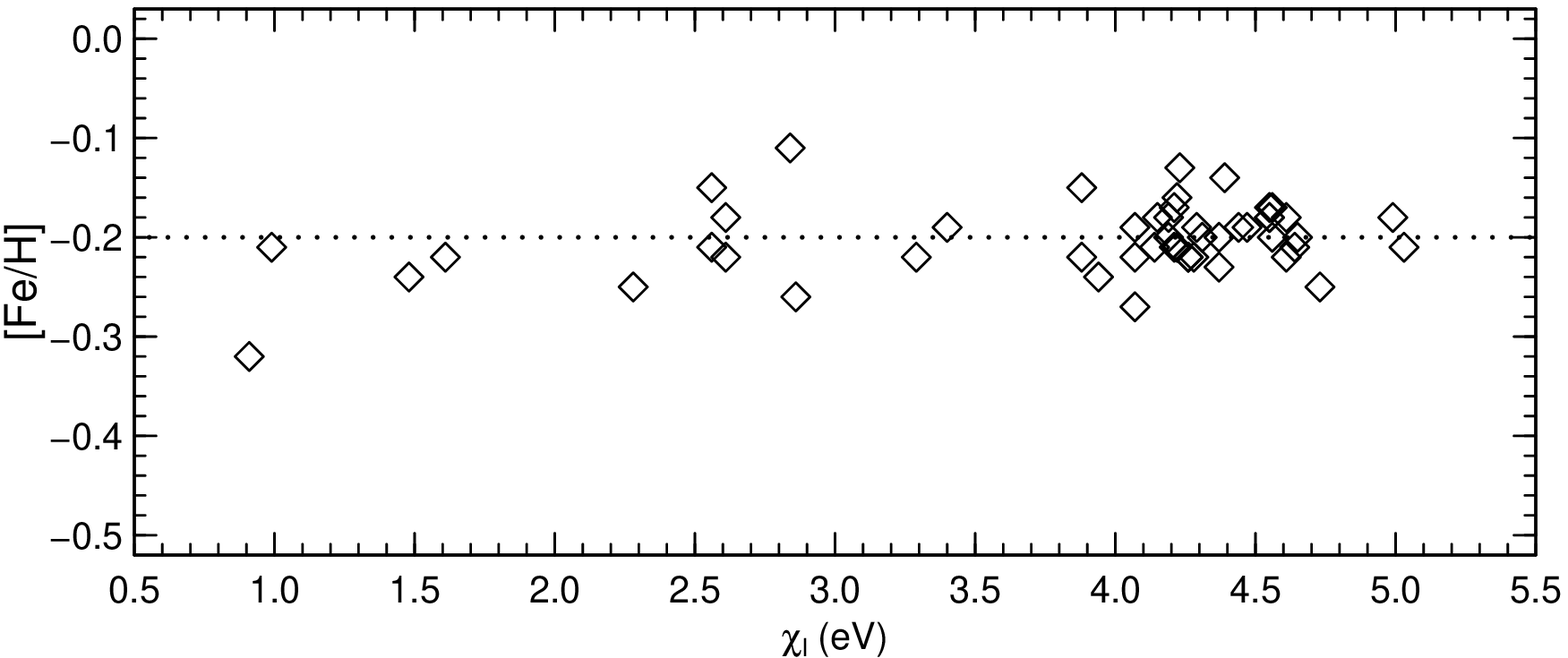,width=120truemm,angle=0,clip=}}
\vspace{-2mm}
\captionb{1}{The [Fe\,{\sc i}/H] abundance values versus the lower
excitation potential $\chi_{\rm l}$.  The mean abundance
([Fe\,{\sc i}/H]$=-0.20$~dex) is shown as a dotted line.}
}
\vspace{3mm}

\vbox{
\centerline{\psfig{figure=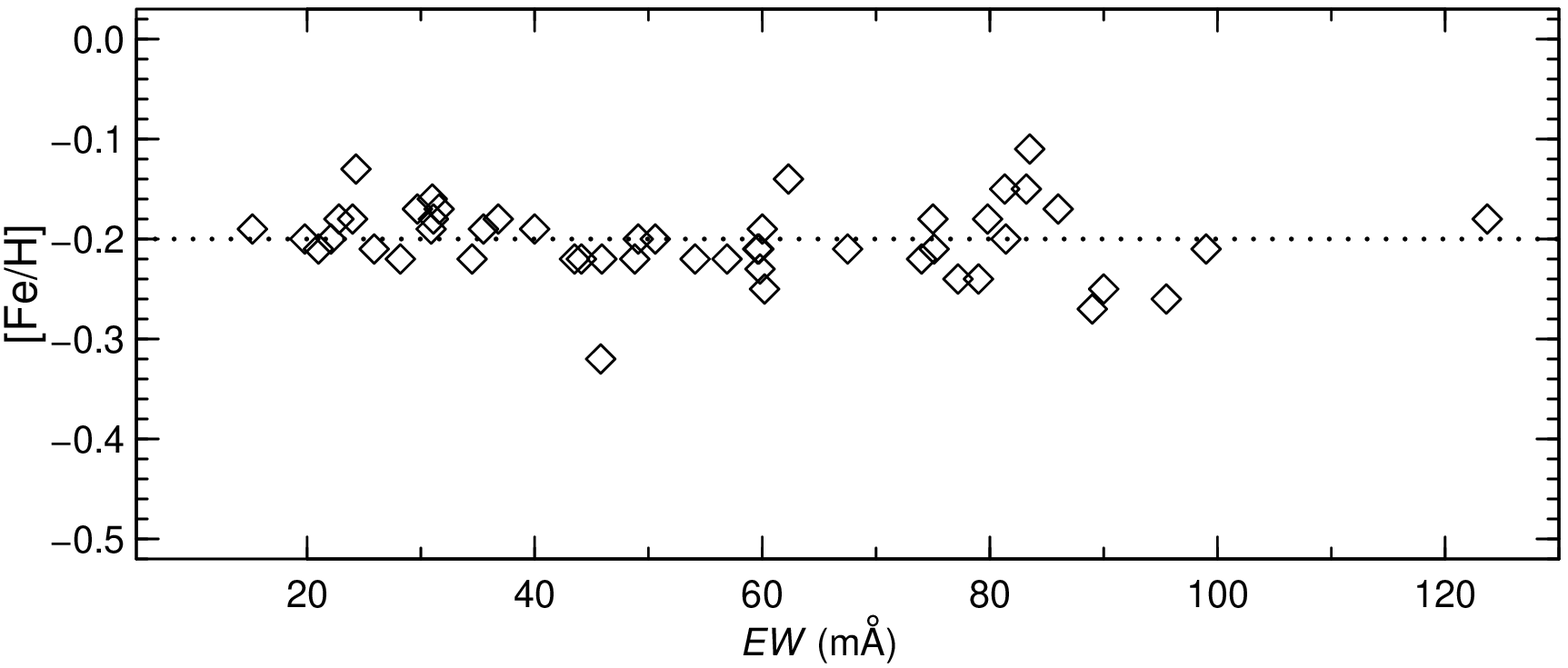,width=120truemm,angle=0,clip=}}
\vspace{-2mm}
\captionb{2}{The [Fe\,{\sc i}/H] abundance values versus
the equivalent widths. The mean abundance ([Fe\,{\sc
i}/H]$=-0.20$~dex) is shown as a dotted line.}
}
\vspace{3mm}

\subsectionb{2.2}{Mass determination}

The mass of 29~Dra was evaluated from its effective temperature,
luminosity and the isochrones from Girardi et al.\ (2000).  The
luminosity $\log\,(L/L_{\odot})=1.44$ was calculated from the {\it
Hipparcos} parallax $\pi=9.66$~mas (van Leeuwen 2007) and $V =
6.64$~mag,
the bolometric correction calculated according to Alonso et al.\
(1999), and the above mentioned $E_{B-V}=0.02$.  The mass of 29~Dra
$\sim$\,1.2\,$M_{\odot}$ was found.

\subsectionb{2.3}{Spectrum syntheses}

The method of synthetic spectra was used to determine carbon abundance
from the C$_2$ line at 5135.5~{\AA}.  The interval 7980--8130~{\AA},
containing strong $^{12}$C$^{14}$N and $^{13}$C$^{14}$N features, was
used for the nitrogen abundance and $^{12}$C/$^{13}$C ratio
determinations.  The $^{12}$C/$^{13}$C ratio was determined
from the (2,0) $^{13}$C$^{12}$N feature at 8004.7~{\AA}.  All
log\,$gf$ values were calibrated to fit to the solar spectrum by Kurucz
(2005) with solar abundances from Grevesse \& Sauval (2000).  In
Figure~3 we show several examples of synthetic spectra in the vicinity
of $^{12}$C$^{14}$N lines.

\vspace{2mm}
\vbox{
\centerline{\psfig{figure=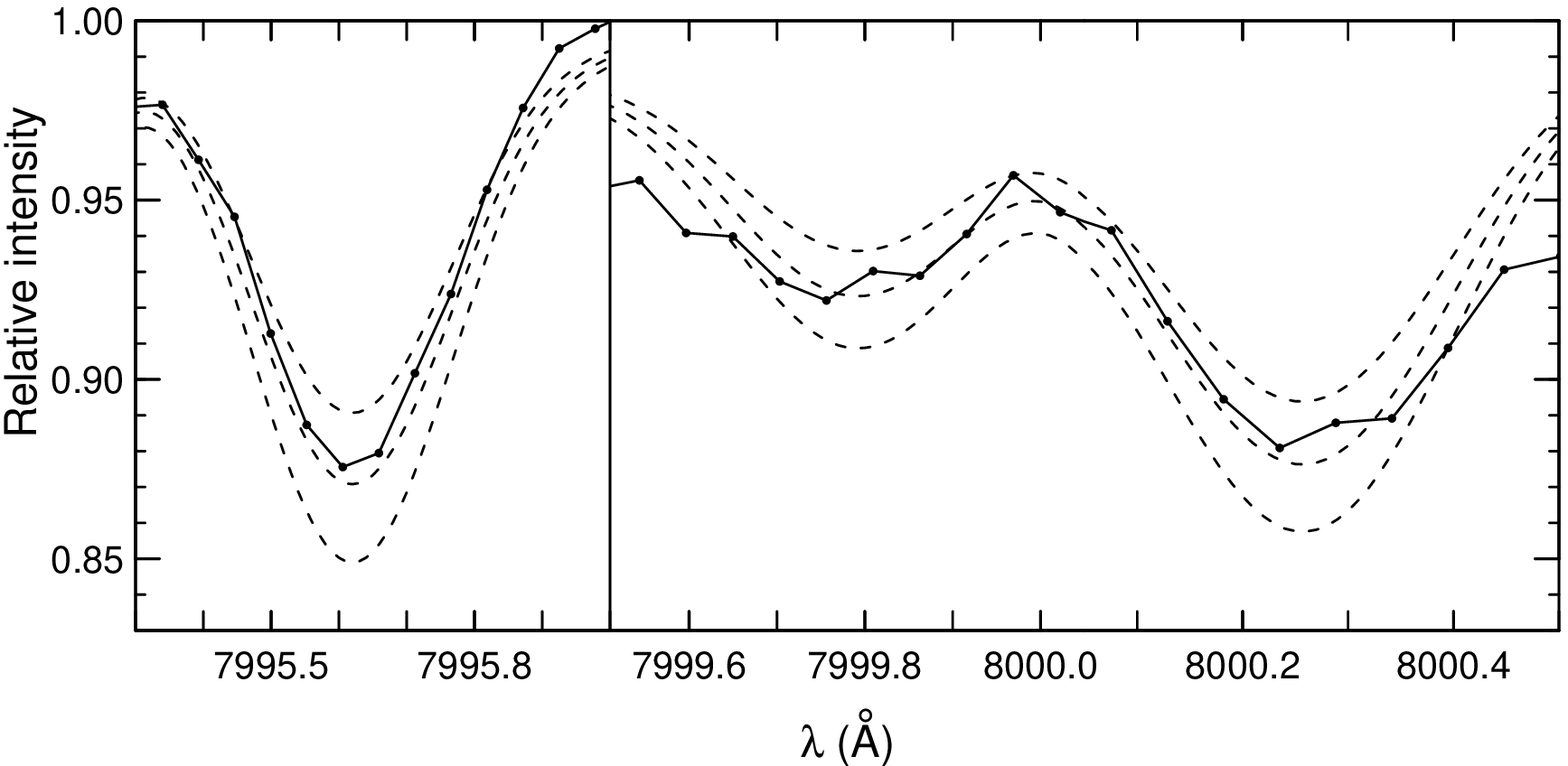,width=125truemm, angle=0,clip=}}
\vspace{-1mm}
\captionb{3}{Synthetic spectrum fits to $^{12}$CN lines.
The observed spectrum is shown as a solid line. The dashed lines are
synthetic spectra with [N/Fe] = 0.02, 0.12 and 0.22~dex.}
}
\vspace{2mm}

The oxygen abundance was determined from the forbidden [O\,{\sc i}] line
at 6300.31~\AA\ (Figure~4) with the oscillator strength values for
$^{58}$Ni and $^{60}$Ni from Johansson et al.  (2003) and the values
log~$gf = -9.917$ obtained by fitting to the solar spectrum (Kurucz
2005) and log~$A_{\odot}=8.83$ taken from Grevesse \& Sauval (2000).

The abundance of Na\,{\sc i} was estimated using the line 5148.84~\AA\
which due to rotational broadening is blended by Ni\,{\sc i} line at
5148.66~\AA.  These two lines are distinct in the Sun, so we were able
to calibrate their log\,$gf$ values using the solar spectrum.  However,
the sodium abundance value in our study is affected by uncertainty of
the nickel abundance determination, originating from the Equivalent
Widths method.  Fortunately, the line-to-line scatter of [Ni/H]
determination from 24 lines of Ni\,{\sc i} was as small as 0.04~dex.

For the evaluation of Zr\,{\sc i} abundance the lines at 5385.13~\AA,
6127.48~\AA\ and 6134.57~\AA\ were used.  Evaluation of Y\,{\sc ii}
abundance (Figure~5) was based on 5402.78~\AA, Pr\,{\sc ii} on
5259.72~\AA\, La\,{\sc ii} on 6390.48~\AA, Ce\,{\sc ii} on 5274.22~\AA\
and 6043.38~\AA\, and Nd\,{\sc ii} on 5276.86~\AA\ lines.

The abundance of Eu\,{\sc ii} was determined from the 6645.10~\AA\ line
(Figure~6).  The hyperfine structure of Eu\,{\sc ii} was taken into
account when calculating the synthetic spectrum.  The wavelength,
excitation energy and total log~$gf = 0.12$ were taken

\vspace{3mm}
\vbox{
\centerline{\psfig{figure=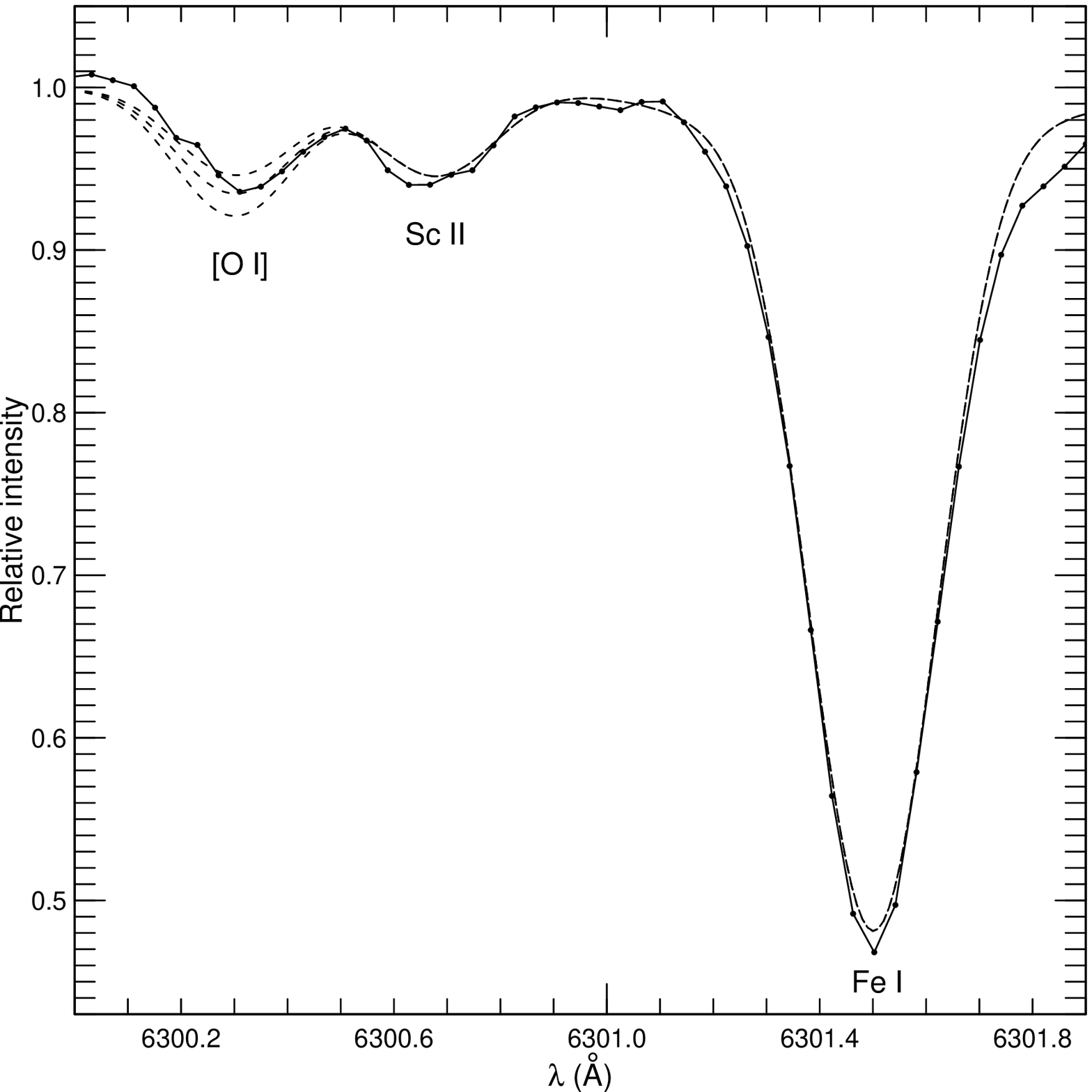,width=95truemm,angle=0,clip=}}
\vspace{-1mm}
 \captionb{4}{Synthetic spectrum fit to forbidden [O\,{\sc i}] line at
6300~\AA. The observed spectrum
is shown as a solid line. The  dashed lines are synthetic spectra with
[O/Fe] = --0.14, --0.04 and 0.06 dex.}
}
\vspace{3mm}

\noindent from Lawler et
al.\ (2001), the isotopic meteoritic fractions of $^{151}{\rm Eu}$,
47.77\%, and $^{153}{\rm Eu}$, 52.23\% and isotopic shifts were taken
from Biehl (1976).

Due to the rotation of RS CVn stars, their spectral lines are broadened,
so it is important to use a correct value of $v\,{\rm sin}\,i$ in the
synthetic spectrum calculations.  We used $v\,{\rm sin}\,i=6.7~{\rm
km\,s}^{-1}$ from Fekel (1997), which fits well the profiles of spectral
lines of 29~Dra.

\subsectionb{2.4}{Estimation of uncertainties}

The sources of uncertainty were described in detail in Paper~I.
The sensitivity of the abundance estimates to changes in the atmospheric
parameters for the assumed errors ($\pm~100$~K for $T_{\rm eff}$, $\pm
0.3$~dex for log~$g$ and $\pm 0.3~{\rm km~s}^{-1}$ for $v_{\rm t}$) is
illustrated in Table~2.  It is seen that possible parameter errors do
not affect the abundances seriously; the element-to-iron ratios, which
we use in our discussion, are even less sensitive.

The scatter of the deduced line abundances $\sigma$, presented in
Table~3, gives an estimate of the uncertainty due to the random errors,
e.g., in the continuum placement and the line parameters (the mean value
of $\sigma$ is 0.05~dex).  Thus the uncertainties in the derived
abundances originating from the random errors are close to this value.

\vspace{3mm}
\vbox{
\centerline{\psfig{figure=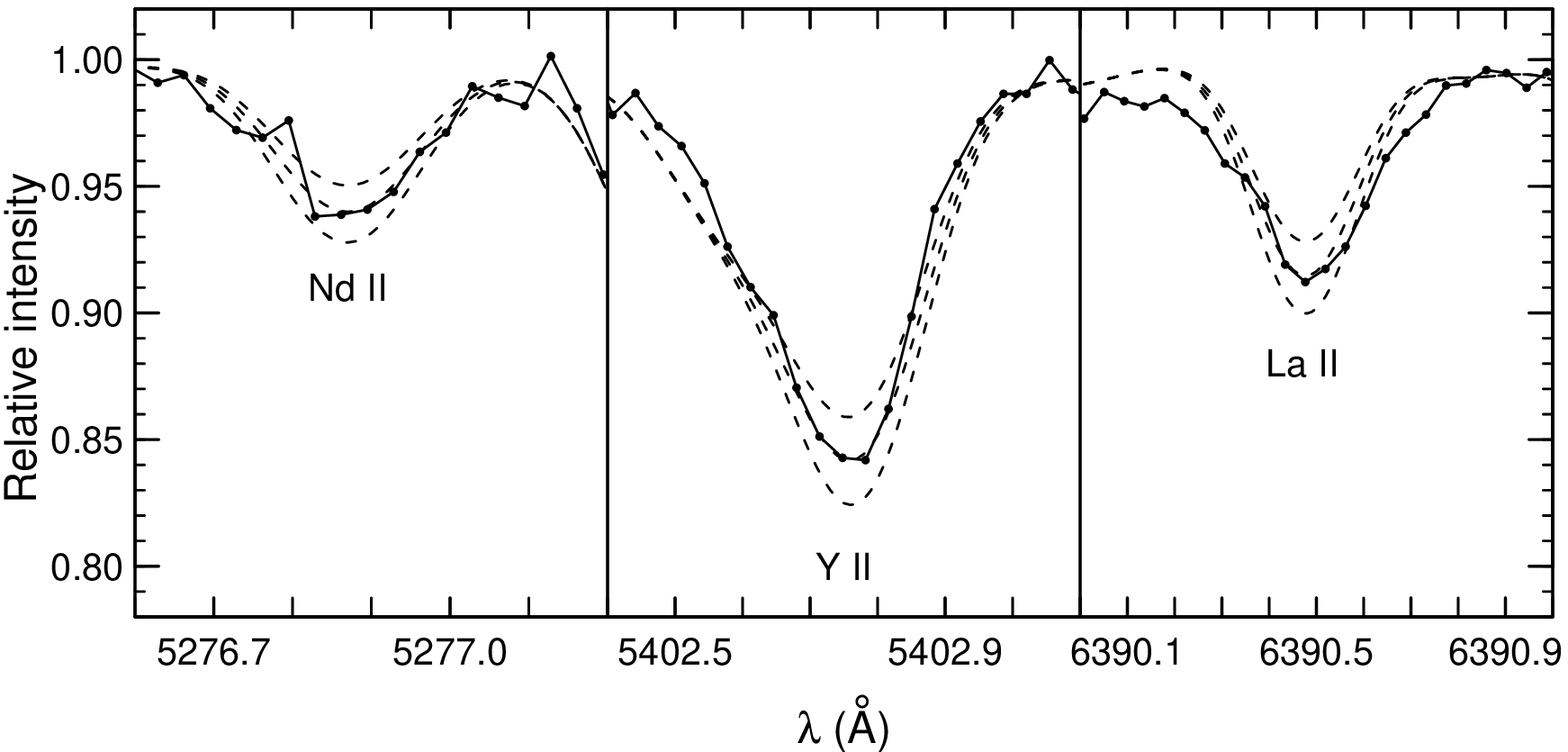,width=125truemm,angle=0,clip=}}
\vspace{-1mm}
\captionb{5}{Synthetic  spectrum fit to the Nd\,{\sc ii} line at
5276.806~\AA,  Y\,{\sc ii} at 5402.78~\AA\  and  La\,{\sc ii} at
6390.48~\AA.  The observed spectrum is shown as a solid
line. The  dashed lines are synthetic spectra  with
${\rm [Nd/Fe]}$ = 0.08, 0.18 and 0.28~dex, ${\rm [Y/Fe]}$ = 0.01, 0.11
and
0.21~dex and ${\rm [La/Fe]}$ = 0.07, 0.17 and 0.27~dex, respectively
for Nd\,{\sc ii}, Y\,{\sc ii} and La\,{\sc ii} lines.}}
\vspace{3mm}

\vspace{3mm}
\vbox{
\centerline{\psfig{figure=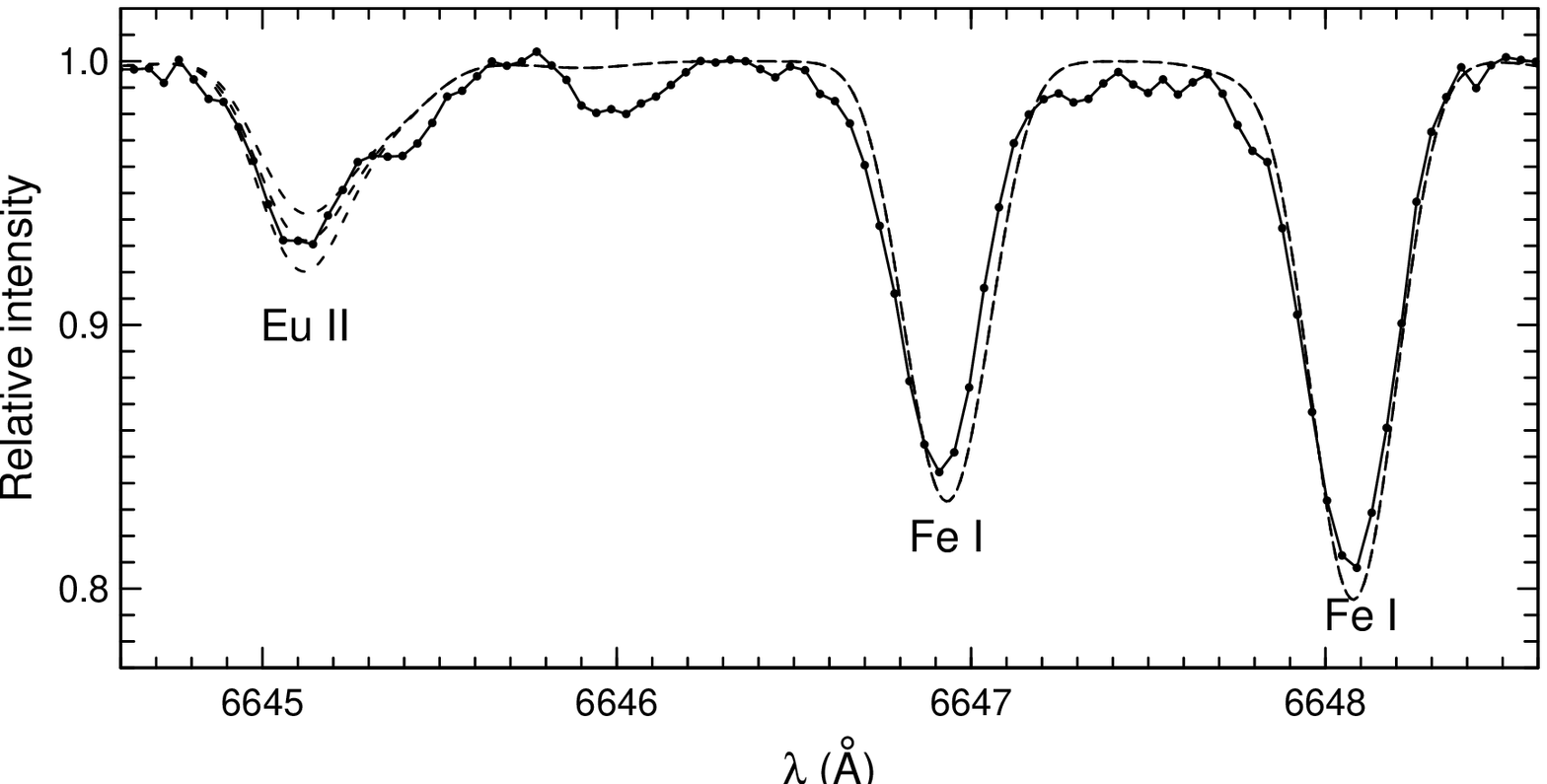,width=125truemm,angle=0,clip=}}
\vspace{-2mm}
\captionb{6}{Synthetic spectrum fit to the Eu\,{\sc ii}  line at
6645.10~\AA\. The observed spectrum
is shown as a solid line. The dashed lines  are synthetic spectra with
${\rm [Eu/Fe]}$ = --0.03, 0.07 and 0.17~dex.}
}
\vspace{3mm}

Since the abundances of C, N and O are bound together by the molecular
equilibrium, we have also investigated how an error in one of them
typically affects the abundance determination of others.

$\Delta{\rm [O/H]}=0.10$ causes
$\Delta{\rm [C/H]}=0.04$ and $\Delta{\rm [N/H]}=0.03$;
$\Delta{\rm [C/H]}=0.10$ causes $\Delta{\rm [N/H]}=-0.12$ and
$\Delta{\rm [O/H]}=0.02$. $\Delta {\rm [N/H]}=0.10$ has no effect
on both carbon and oxygen abundances.

\vspace{3mm}
\begin{center}
\vbox{\small
\tabcolsep=5pt
\begin{footnotesize}
\begin{tabular}{lrrclrrc}
\multicolumn{8}{l}{\parbox{120mm}{
{\small \bf \ \ Table 2.~}{\small Sensitivity
of the abundances to changes of the atmospheric parameters.
The table  entries show the effects on the logarithmic abundance
relative to hydrogen, $\Delta$\,[A/H].
}}}\\
\tablerule
\multicolumn{1}{l}{Ion}&
\multicolumn{1}{c}{$\Delta T_{\rm eff}$}&
\multicolumn{1}{c}{$\Delta {\rm log}~g$}&
\multicolumn{1}{c}{$\Delta v_{\rm t}$}&

\multicolumn{1}{l}{Ion }&
\multicolumn{1}{c}{$\Delta T_{\rm eff}$}&
\multicolumn{1}{c}{$\Delta {\rm log}~g$}&
\multicolumn{1}{c}{$\Delta v_{\rm t}$}\\

\multicolumn{1}{c}{}&
\multicolumn{1}{c}{$+100$~K}&
\multicolumn{1}{c}{$+0.3$}&
\multicolumn{1}{c}{$+0.3 {\rm km~s}^{-1}$}&

\multicolumn{1}{c}{}&
\multicolumn{1}{c}{$+100$~K}&
\multicolumn{1}{c}{$+0.3$}&
\multicolumn{1}{c}{$+0.3 {\rm km~s}^{-1}$}\\
\tablerule
C(C$_2$)		&$	0.00	$ & $	0.11	$ & $	0.00	$ &	Fe\,{\sc ii}	&$	-0.09	$ & $	0.18	$ & $	-0.06	$\\
N(CN)			&$	0.06	$ & $	0.10	$ & $	-0.01	$ &	Co\,{\sc i}		&$	0.06	$ & $	0.06	$ & $	-0.06	$\\
O([O\,{\sc i}])	&$	0.00	$ & $	0.16	$ & $	-0.01	$ &	Ni\,{\sc i}		&$	0.03	$ & $	0.08	$ & $	-0.09	$\\
Na\,{\sc i}		&$	0.12	$ & $	0.00	$ & $	0.00	$ &	Y\,{\sc ii}		&$	0.00	$ & $	0.13	$ & $	-0.04	$\\
Si\,{\sc i}		&$	-0.03	$ & $	0.07	$ & $	-0.04	$ &	Zr\,{\sc i}		&$	0.19	$ & $	0.00	$ & $	-0.01	$\\
Ca\,{\sc i}		&$	0.09	$ & $  -0.01	$ & $	-0.09	$ &	La\,{\sc ii}	&$	0.01	$ & $	0.24	$ & $	0.08	$\\
Sc\,{\sc ii}	&$	-0.01	$ & $	0.14	$ & $	-0.08	$ &	Ce\,{\sc ii}	&$	0.01	$ & $	0.14	$ & $	-0.04	$\\
Ti\,{\sc i}		&$	0.14	$ & $	0.01	$ & $	-0.06	$ &	Pr\,{\sc ii}	&$	0.02	$ & $	0.14	$ & $	-0.02	$\\
V\,{\sc i}		&$	0.16	$ & $	0.01	$ & $	-0.08	$ &	Nd\,{\sc ii}	&$	0.01	$ & $	0.14	$ & $	-0.01	$\\
Cr\,{\sc i}		&$	0.08	$ & $	-0.01	$ & $	-0.07	$ &	Eu\,{\sc ii}	&$	-0.01	$ & $	0.16	$ & $	-0.01	$\\
Fe\,{\sc i}		&$	0.05	$ & $	0.04	$ & $	-0.07	$ &		&		&		&		\\
\\
C/N &	$	-0.31	$ & $	0.05	$ & $	0.05	$ & \textsuperscript{12}C/\textsuperscript{13}C & $	0	$ & $	1	$ & $	0	$  \\
\tablerule
\end{tabular}
\end{footnotesize}
}
\end{center}

\vspace{3mm}

\begin{center}
\vbox{\small
\tabcolsep=10pt
\begin{footnotesize}
\begin{tabular}{lrrclrrc}
\multicolumn{8}{l}{\parbox{110mm}{
 {\small \bf \ \ Table 3.}{\small \ Element abundances relative to
hydrogen, [A/H]. $\sigma$ is a standard deviation in the
mean value due to the line-to-line scatter within the species. $N$ is
the number of lines used for the abundance determination.
 }}}\\
\tablerule
\multicolumn{1}{l}{Ion }&
\multicolumn{1}{c}{$N$}&
\multicolumn{1}{c}{[A/H]}&
\multicolumn{1}{c}{$\sigma$ }&
\multicolumn{1}{l}{Ion }&
\multicolumn{1}{c}{$N$}&
\multicolumn{1}{c}{[A/H]}&
\multicolumn{1}{c}{$\rm \sigma$ } \\
\tablerule
C(C$_2$)	& 1	& $-0.34$	& $-$  	& Fe\,{\sc ii}	& 6	& $-0.21$	& $0.07$ \\
N(CN)	& 4	& $-0.12$	& $0.01$  	& Co\,{\sc i}	& 8	& $-0.13$	& $0.07$ \\
O([O\,{\sc i}])	& 1	& $-0.24$	& $-$  	& Ni\,{\sc i}	& 24	& $-0.22$	& $0.04$ \\
Na\,{\sc i}	& 1	& $-0.02$	& $-$  	& Y\,{\sc ii}	& 1	& $-0.09$	& $-$ \\
Si\,{\sc i}	& 7	& $-0.11$	& $0.12$  	& Zr\,{\sc i}	& 3	& $-0.18$	& $0.08$ \\
Ca\,{\sc i}	& 4	& $0.04$	& $0.07$  	& La\,{\sc ii}	& 1	& $-0.03$	& $-$ \\
Sc\,{\sc ii}	& 5	& $-0.14$	& $0.06$  	& Ce\,{\sc ii}	& 2	& $-0.30$	& $0.01$ \\
Ti\,{\sc i}	& 18	& $-0.12$	& $0.06$  	& Pr\,{\sc ii}	& 1	& $-0.24$	& $-$ \\
V\,{\sc i}	& 19	& $-0.06$	& $0.06$  	& Nd\,{\sc ii}	& 1	& $-0.02$	& $-$ \\
Cr\,{\sc i}	& 7	& $-0.10$	& $0.07$  	& Eu\,{\sc ii}	& 1	& $-0.13$	& $-$ \\
Fe\,{\sc i}	& 50	& $-0.20$	& $0.04$  	& 	&	&	&	\\
\tablerule
\end{tabular}
\end{footnotesize}
}
\end{center}

\sectionb{3}{RESULTS AND DISCUSSION}

As a result, for 29~Dra we found the following atmospheric
parameters:  $T_{\rm eff}=4720$~K, ${\rm log}\,g=2.5$, $v_{\rm
t}=1.4~{\rm km}\,s^{-1},$ ${\rm [Fe/H]}=-0.20$, ${\rm [C/Fe]} = -0.14$,
${\rm [N/Fe]} = 0.08$, ${\rm [O/Fe]} = -0.04$, as well as the ratios C/N
= 2.40 and $^{12}$C/$^{13}$C = 16.  The element abundances [A/H] and the
$\sigma$ values (the line-to-line scatter) are listed in Table~3 and
displayed in Figure~7.

The effective temperature and surface gravity values for 29~Dra given in
different papers are quite similar.  T$_{\rm eff}=4700$~K was determined
by Randich et al.\ (1994), 4800~K by Berdyugina (1994) and Za\v{c}s et
al.\ (1997) and 4720~K in the present paper.  Our value of log~$g=2.5$
is in agreement with the result of Za\v{c}s et al.\ (1997).  Randich et
al.\ (1994) have found the value larger by 0.1~dex.  From the wings of
MgH lines, Berdyugina (1994) has found log~$g=2.7$.

The [Fe/H] value of $-0.20$~dex determined in the present paper is close
to the value of Za\v{c}s et al., $-0.24$~dex.  A slightly higher value,
$-0.05$~dex, was found by Berdyugina and an exceptionally low value of
$-0.8$~dex by Randich et al.

In Figure~7, the element to iron ratios for 29~Dra are compared with the
results of other investigations.  The element to iron ratios for the
iron peak, $s$- and $r$-process elements are solar.  Abundances of
$\alpha$-elements are enhanced by about 0.1~dex.

\vspace{3mm}
\vbox{
\centerline{\psfig{figure=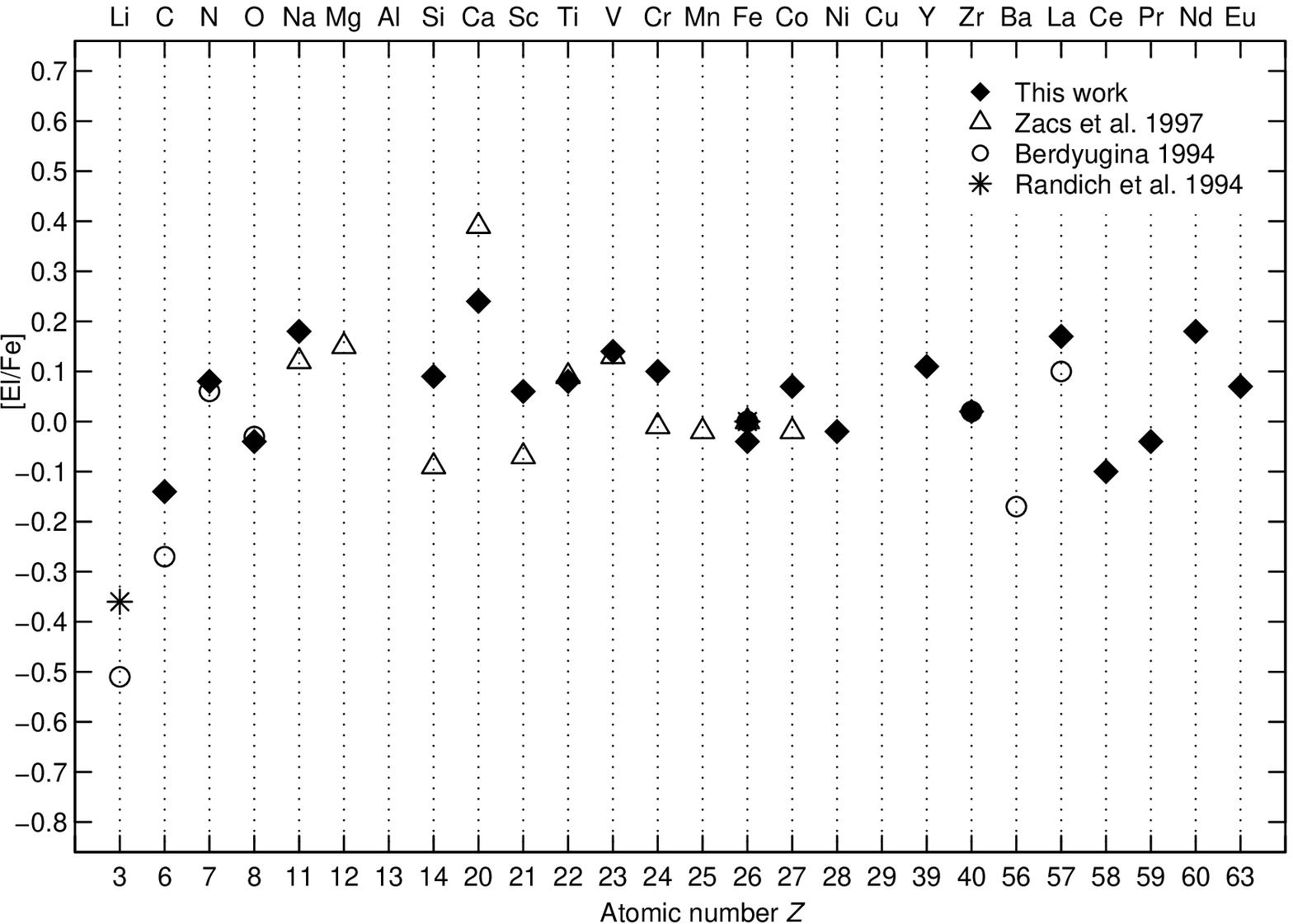,width=120mm, angle=0,clip=}}
\vspace{-1mm}
\captionb{7}{Element abundances for 29~Dra, as determined in this work
(filled diamonds), by Za\v{c}s et al.\
(1997, triangles), Berdyugina (1994, circles) and Randich et al.\ (1994,
asterisks).}
}
\vspace{3mm}

\subsectionb{3.1}{The ratios C/N and $^{12}$C/$^{13}$C}

The evolutionary sequences in the luminosity versus effective
temperature diagram by Girardi et al.\ (2000) show that 29~Dra with its
luminosity ${\rm log}(L/L_{\odot})=1.44$ is a first ascent giant lying
slightly below the red giant sequence bump at ${\rm
log}\,(L/L_{\odot})=1.6$ (Charbonnel \& Lagarde 2010).  According to the
mentioned model of mixing, carbon and nitrogen abundances in 29~Dra
should be altered only by the first dredge-up.  The ratio
$^{12}$C/$^{13}$C for the first ascent giants with the mass of
29~Dra
(1.2\,$M_{\odot}$) should be lowered to the value of about 26
(Charbonnel \& Lagarde 2010).  However, the ratio found in this paper
is 16.

In Figure~8, we compare C/N and $^{12}$C/$^{13}$C ratios of 29~Dra with
the theoretical models including extra mixing.  The model, called `cool
bottom processing' (CBP), was proposed by Boothroyd \& Sackmann (1999)
and the model, called `thermohaline mixing' (TH), was proposed by
Charbonnel \& Lagarde (2010).  The position of 29~Dra in Figure~8
indicates that its ratio of carbon isotopes has been altered by extra
mixing.  The low value of the $^{12}$C/$^{13}$C ratio was also found in
$\lambda$~And (Paper~I).

\vspace{3mm}
\vbox{
\centerline{\psfig{figure=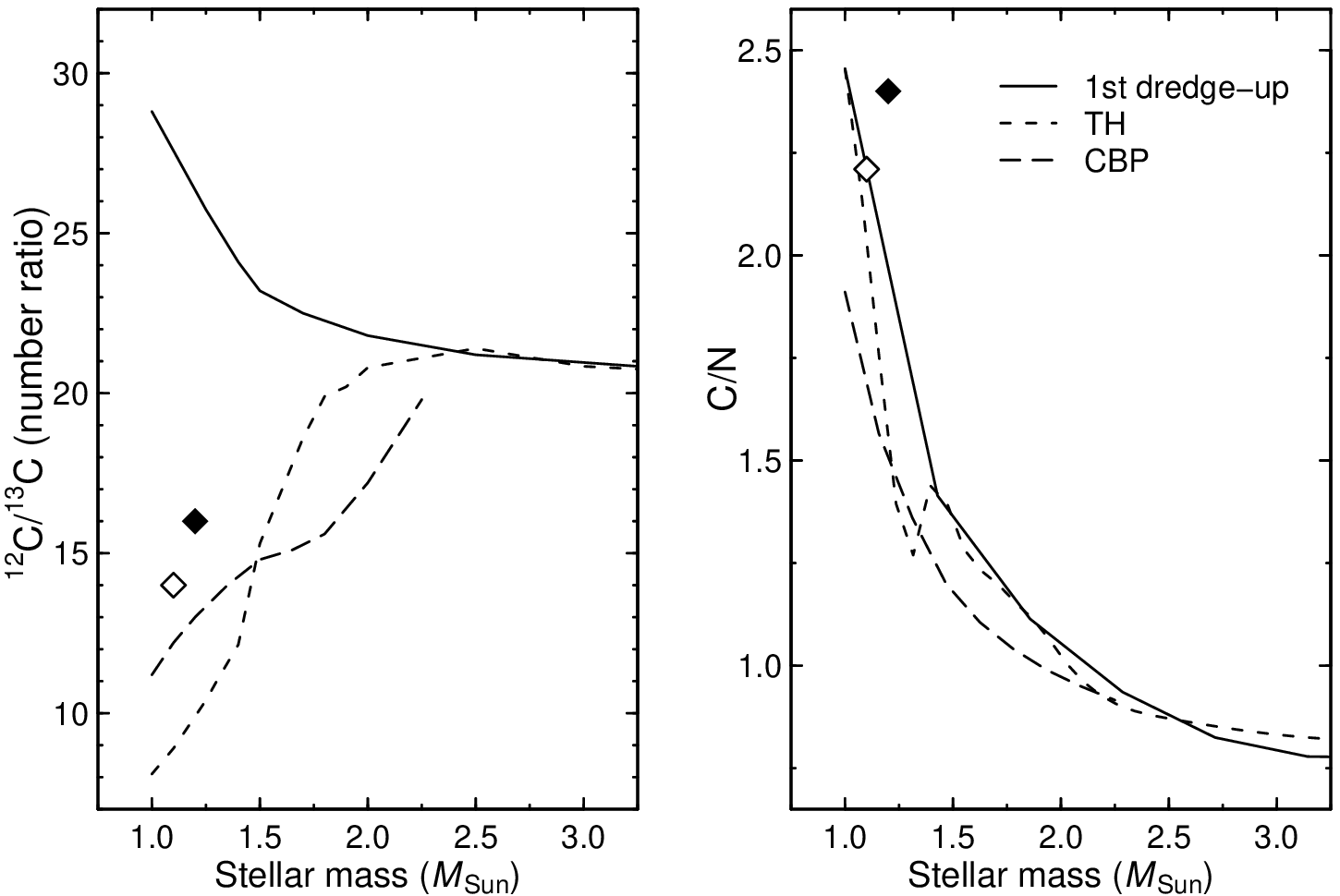,width=120truemm,angle=0,clip=}}
\vspace{-1mm}
\captionb{8}{Comparisons of C/N and $^{12}$C/$^{13}$C ratios in
29~Dra (filled diamonds) and  $\lambda$~And (open diamonds, Paper~I)
with the theoretical predictions explained in the text.}
}
\vspace{3mm}

The abundance of lithium is also very sensitive to mixing.
During
the first dredge-up, for a star of the mass of 29 Dra, Li abundance
drops to log\,$A{\rm (Li)}=1.45$ (Charbonnel \& Lagarde
2010).  However, the available determinations of Li abundance in 29~Dra
show lower abundances.  Berdyugina (1994) finds log\,$A{\rm (Li)}=0.6$
and Randich et al.\ (1994) find log\,$A{\rm (Li)}\le 0$.  The low
abundance of Li is compatible with the thermohaline extra-mixing
model by Charbonnel \& Lagarde (2010), which predicts the lowering of Li
abundance to the value of about 0.07~dex.  The observed
Li abundances in $\lambda$~And are even lower (Savanov \& Berdyugina
1994; Randich et al.\ 1994; Mallik 1998).

Thus, both $\lambda$~And and 29~Dra, investigated by now in our
program, give a hint that extra-mixing processes may start acting in
these low-mass chromospherically active fast rotating stars slightly
earlier than at the bump of the red giant sequence in non-active
stars.

\thanks{This project has been supported by the European Commission
through the Baltic Grid II project.}

\References

\refb Abt~H.~A., Biggs.~E.~S. 1972, Bibliography of Stellar Radial
Velocities, KPNO, Tucson

\refb Alonso~A., Arribas~S., Mart\'{i}nez-Roger~C. 1999, A\&AS, 140, 261

\refb Berdyugina S.~V. 1994, Astronomy Letters, 20, 796

\refb Bidelman W. P. 1954, ApJS, 1, 175

\refb Biehl~D. 1976, Diplomarbeit, Christian-Albrechts-Universit\"at
Kiel, Institut f\"ur Theoretische Physik und Sternwarte

\refb Boothroyd~A.~I., Sackman~I.~J. 1999, ApJ, 510, 232

\refb Charbonnel C., Lagarde N. 2010, A\&A, 522, A10

\refb Fekel~F.~C. 1997, PASP, 109, 514

\refb Fekel~F.~C., Simon~T. 1985, AJ, 90, 812

\refb Fekel~F.~C., Henry~G.~W., Busby~M.~R., Eitter~J. 1993, AJ, 106,
2370

\refb Girardi L., Bressan A., Bertelli G., Chiosi C. 2000, A\&AS, 141,
371

\refb Grevesse N., Sauval A. J. 2000, in {\it Origin of Elements in the
Solar System, Implications of Post-1957 Observations}, ed. O. Manuel,
Kluwer, p.\,261

\refb Hakkila~J., Myers~J.~M., Stidham~B.~J., Hartmann~D.~H. 1994, AJ,
114, 2043

\refb Hall~D.~S., Henry~G.~W., Louth~H., Renner~T.~R., Shore~S.~N. 1982,
IBVS, 2109, 1

\refb Hauck~B., Mermilliod~M. 1998, A\&AS, 129, 431

\refb Johansson~S., Litzen~U., Lundberg~H., Zhang~Z. 2003, ApJ , 584,
107

\refb Kurucz~R. L. 2005, {\it New Atlases for Solar Flux, Irradiance,
Central Intensity, and Limb Intensity}, Mem. SA Ital. Suppl., 8, 189

\refb Lawler J. E., Wickliffe M. E., Den Hartog E. A. 2001, ApJ, 563,
1075

\refb van Leeuwen~F. 2007, {\it Hipparcos, the New Reduction of the Raw
Data}, Astrophysics and Space Science Library, vol. 350

\refb Mallik S. V. 1998, A\&A, 338, 623

\refb Randich~S., Giampapa~M.~S., Pallavicini~R. 1994, A\&A, 283, 893

\refb Roman~N.~G. 1955, ApJS, 2, 195

\refb Savanov~I.~S., Berdyugina~S.~V. 1994, Astronomy Letters, 20, 227

\refb Tautvai\v{s}ien\.{e}~G., Barisevi\v{c}ius~G., Berdyugina~S.,
Chorniy~Y., Ilyin~I. 2010, Baltic Astronomy, 19, 95 (Paper~I)

\refb Za\v{c}s~L., Musaev~F.~A., Bikmaev~I.~F., Alksnis~O. 1997,
A\&AS, 122, 31

\refb Zboril~M., Messina~S. 2009, AN, 330, 377

\end{document}